\documentclass[aps,prd,showpacs,superscriptaddress,preprintnumbers,dblfloatfix,floatfix,nofootinbib,twocolumn]{revtex4-2}
\usepackage{graphicx}

\usepackage{amsmath,amssymb}
\usepackage{epsfig}
\usepackage{booktabs}
\usepackage{hyperref}
\usepackage{color}
\usepackage{placeins}
\usepackage{amsfonts}
\usepackage{dsfont}
\usepackage[dvipsnames]{xcolor}
\usepackage{appendix}
\usepackage[export]{adjustbox}
\usepackage{multirow}
\usepackage[capitalize]{cleveref}
\usepackage[font={small}]{caption}   
\usepackage[normalem]{ulem}

\definecolor{myred}{rgb}{0.7, 0, 0}
\definecolor{myblue}{rgb}{0, 0, 0.7}
\definecolor{mygreen}{rgb}{0.04, 0.7, 0.5}
\definecolor{mygray}{rgb}{0.1, 0.1, 0.1}
\hypersetup{colorlinks,citecolor=myred,linkcolor=myblue,urlcolor=myblue,linktocpage=true}

\newcommand{\skipnew}[1]{}

\numberwithin{equation}{section}
\renewcommand{\theequation}{\arabic{section}.\arabic{equation}}

\newcommand{\SO}{\operatorname{SO}}
\newcommand{\CP}{\mathrm{CP}}

\newcommand{\hc}{\mathrm{h.c.}}
\newcommand{\Tr}{\mathrm{Tr}}
\newcommand{\SU}{\mathrm{SU}}
\newcommand{\U}{\mathrm{U}}

\newcommand{\Sym}{\mathrm{Sym}}

\def\UMD{\small{Maryland Center for Fundamental Physics, Department of Physics, \\ University of Maryland, College Park, MD 20742, USA}}

\usepackage{tikz,xcolor,hyperref}
\definecolor{lime}{HTML}{A6CE39}
\DeclareRobustCommand{\orcidicon}{%
	\begin{tikzpicture}
	\draw[lime, fill=lime] (0,0) 
	circle [radius=0.16] 
	node[white] {{\fontfamily{qag}\selectfont \tiny ID}};	\draw[white, fill=white] (-0.0625,0.095) 
	circle [radius=0.007];	\end{tikzpicture}
	\hspace{-2mm}}
\foreach \x in {A,...,Z}{%
	\expandafter\xdef\csname orcid\x\endcsname{\noexpand\href{https://orcid.org/\csname orcidauthor\x\endcsname}{\noexpand\orcidicon}}
}

\definecolor{deepred}{RGB}{150,0,0}

\begin{document}
\title{A spurionic interpretation of RG-stable parameter relations in the two-Higgs-doublet model}
\author{Zackaria~Chacko\orcidB{}\,}
\email{zchacko@umd.edu}
\affiliation{\UMD}
\author{Shourya~Mukherjee\orcidC{}\,}
\email{mshourya@umd.edu}
\affiliation{\UMD}

\date{\today}

\begin{abstract}
Recent work on the two-Higgs-doublet model (2HDM) has uncovered basis-invariant relations among the parameters in the scalar potential that are stable under renormalization group (RG) evolution but do not coincide with any of the standard Higgs-family or Generalized-$\CP$ symmetry classes of the scalar sector. These relations have subsequently been reformulated in terms of so-called GOOFy transformations. In this paper, we show that the same fixed surface admits a conventional interpretation in terms of spurions. The gauge-covariant kinetic terms of the 2HDM possess a Higgs-family $\U(2)_H$ symmetry, and the couplings in the potential may be organized into irreducible spurions under its nontrivial $\SU(2)_H$ subgroup. We first explain the one-loop closure in the spurion language and then establish stability to all orders in the scalar and gauge couplings in any mass-independent scheme. General Yukawa couplings introduce additional Higgs-family covariants and can spoil the scalar-sector selection rule. We therefore consider the six conventional Higgs-family and Generalized-$\CP$ symmetries of the 2HDM, imposed as exact symmetries of the full Lagrangian, and show that closure is controlled by whether the relevant Higgs-family triplet contains an invariant direction. The spurion formulation thus provides a basis-covariant and conventional group-theoretic explanation of the RG stability of the 2HDM parameter relations.
\end{abstract}

\maketitle

\section{Introduction}

A familiar lesson of quantum field theory is that relations among couplings are naturally preserved by renormalization group (RG) evolution when they are enforced by a symmetry~\cite{tHooft:1979rat}. Conversely, in a general theory, one expects relations that are not protected by a symmetry to be destabilized by radiative corrections. The two-Higgs-doublet model (2HDM) has recently provided a striking setting in which RG-stable relations arise without appearing to correspond to any of the conventional symmetries of the scalar potential.

The 2HDM extends the Standard Model (SM) scalar sector by introducing a second $\SU(2)_L$ doublet that has identical gauge quantum numbers. The classification of its scalar potential under Higgs-family and Generalized-$\CP$ transformations, together with its basis-covariant formulation, is well established~\cite{Ivanov:2005hg,Ivanov:2006yq,Ferreira:2009wh,Ferreira:2020ana,Bento:2020jei}. Ref.~\cite{Ferreira:2023dke} identified a set of basis-invariant parameter relations which do not coincide with any of these conventional symmetry classes, but nevertheless define an RG-invariant surface in the scalar theory with gauge and quartic interactions. This unexpected stability has raised the question of whether there is a deeper underlying principle that protects these relations against radiative corrections. 

Several complementary explanations have since been proposed. Ref.~\cite{Ferreira:2023dke} formulated the relations in terms of ``GOOFy'' transformations, which leave the relevant parameter structures invariant but act non-trivially on the canonical kinetic terms and therefore are not conventional symmetries of the full Lagrangian. Generalizations and systematic constructions of GOOFy transformations were subsequently developed in Refs.~\cite{Trautner:2025yxz, Grzadkowski:2026gkx}. Conditions for the one-loop effective potential to be invariant under GOOFy transformations were discussed in Ref.~\cite{Ferreira:2025ate}. The extension of GOOFy transformations to the full SM has been considered in Refs.~\cite{deBoer:2025jhc,Trautner:2025prm}, and the generalization to theories with larger numbers of Higgs doublets has been explored in Ref.~\cite{deMedeirosVarzielas:2026jsk}. 

Alternative explanations for the RG stability were considered in Refs.~\cite{Haber:2025cbb,deBoer:2026ktb,Doring:2024kdg,Pilaftsis:2026wyw}. Ref.~\cite{Haber:2025cbb} studied an analogous RG-stable relation in a simpler scalar theory and showed that it can be understood as an ordinary symmetry after complexification of the field space. Refs.~\cite{deBoer:2026ktb,Doring:2024kdg} showed that outer automorphisms of the symmetry structure provide sufficient conditions for RG-invariant parameter hyperplanes, thereby giving a symmetry-theoretic criterion that need not correspond to an ordinary symmetry of the action. Ref.~\cite{Pilaftsis:2026wyw} instead developed a constructive spurion approach in which the breaking of classical scale symmetry is itself encoded by dimensionful spurions, together with spurions that correspond to the breaking of additional global symmetries. Their transformation properties constrain the allowed structure of the beta functions, allowing the identification of RG-invariant relations among the parameters.

In this paper, we propose an alternative approach to the problem that is directly tied to the Higgs-family structure of the 2HDM. In this framework, RG closure is formulated as a representation-theoretic statement in the physical coupling space rather than through automorphisms or scale-invariant directions. Our proposal is therefore complementary to these earlier works, differing in both construction and emphasis. 

Our approach is based on a direct formulation of the 2HDM fixed surface in terms of irreducible spurions of the Higgs-family group. Since the two Higgs doublets carry identical SM gauge quantum numbers, their gauge-covariant kinetic terms respect a $\U(2)_H$ basis symmetry. Under its non-trivial $\SU(2)_H$ factor, the scalar couplings decompose into singlet, triplet, and quintet spurions. In this language, the fixed surface is simply the locus on which the singlet component of the quadratic coupling and the triplet component of the quartic coupling vanish. The question of RG closure then becomes purely representation-theoretic -- can the remaining dimensionless spurions generate a covariant that transforms as a triplet? We show that they cannot and prove the resulting closure to all orders in the scalar-gauge sector using a character-theory generating function. In this way, the RG-stable relations follow directly from conventional spurion selection rules in the physical 2HDM coupling space.

The connection with 't~Hooft naturalness~\cite{tHooft:1979rat} should therefore be understood at the level of symmetry selection rules. In the usual argument, symmetry forbids the radiative generation of a symmetry-breaking coupling. In our case, once the couplings have been promoted to spurions, the theory is formally invariant under the $\SU(2)_H$ symmetry, with the couplings transforming as specific representations of $\SU(2)_H$.

The fixed surface is identified with the vanishing of certain specific couplings that have definite transformation properties under $\SU(2)_H$. Then, when these couplings are set to zero, no covariants with the required combination of $\SU(2)_H$ quantum numbers and mass dimension to contribute to their beta functions can be constructed from the remaining couplings.

Before treating the full 2HDM, we illustrate this mechanism in the simpler scalar model considered in Ref.~\cite{Haber:2025cbb}. For this case, the relevant spurion selection rule reduces to charge conservation under a $\U(1)$ symmetry of the kinetic term, making the essential logic particularly transparent. The 2HDM argument is the corresponding non-Abelian generalization.

The inclusion of Yukawa interactions changes the problem qualitatively. General Yukawa couplings carry additional Higgs-family quantum numbers and generate new non-singlet covariants, so the scalar-sector selection rule by itself no longer guarantees the RG stability of the fixed relations. Rather than searching for texture-dependent cancellations, we ask a symmetry-protected question. We impose one of the conventional 2HDM symmetry classes, denoted by $G$, as an exact symmetry of the full Lagrangian, including the Yukawa sector, and ask whether the fixed relations remain preserved under RG evolution. The relevant criterion is whether the real Higgs-family triplet contains a $G$-invariant direction. If no such direction exists, neither scalar nor Yukawa counterterms can generate a triplet coupling. Applying this criterion to the six conventional symmetry classes, we find that $\SO(3)_H$, $\CP2$, and $\CP3$ do not allow any nonvanishing triplet invariant and therefore preserve the fixed relations to all perturbative orders. By contrast, $\mathbb Z_2$, $\U(1)_{PQ}$, and $\CP1$ retain nonzero invariant triplet directions, and the Yukawa sector supplies explicit covariants along those directions. These symmetries therefore do not protect the full set of RG-stable relations.

This paper is organized as follows. Section~\ref{sec:lagrangian} reviews the 2HDM scalar potential and develops the $\SU(2)_H$ spurion decomposition, identifying the fixed surface in basis-invariant form. Section~\ref{sec:toymodel} presents the spurion argument for the simpler scalar model of Ref.~\cite{Haber:2025cbb}. Section~\ref{sec:firstorder} gives the leading-order proof for the 2HDM using representation theory, while Section~\ref{sec:allorders} extends the argument to all orders in the scalar-gauge sector. Section~\ref{sec:yukawa} develops the Yukawa spurion analysis and identifies the symmetry classes under which the scalar surface remains closed to all orders. We summarize our conclusions in Section~\ref{sec:conclusions}. 

\section{The 2HDM spurion decomposition} \label{sec:lagrangian}

\subsection{Higgs-family symmetry}

The most general renormalizable, gauge-invariant 2HDM scalar potential is~\cite{Davidson:2005cw,Haber:2006ue,Ferreira:2009wh,Bento:2020jei}

\newpage

\begin{widetext}
\begin{align}
    V &=\;
    m_{11}^2(H_1^\dagger H_1) + m_{22}^2(H_2^\dagger H_2) - \bigl[m_{12}^2(H_1^\dagger H_2) + \hc\bigr] +\frac{\lambda_1}{2}(H_1^\dagger H_1)^2 +\frac{\lambda_2}{2}(H_2^\dagger H_2)^2 
    \\
    &+\lambda_3(H_1^\dagger H_1)(H_2^\dagger H_2) +\lambda_4(H_1^\dagger H_2)(H_2^\dagger H_1) +\Bigl[\frac{\lambda_5}{2}(H_1^\dagger H_2)^2 +\lambda_6(H_1^\dagger H_1)(H_1^\dagger H_2) +\lambda_7(H_2^\dagger H_2)(H_1^\dagger H_2) +\hc\Bigr]. \nonumber
\label{eq:potential}
\end{align} 
\end{widetext}
Hermiticity implies that $m_{11}^2,m_{22}^2,\lambda_{1,2,3,4}$ are real, while $m_{12}^2$ and $\lambda_{5,6,7}$ may be complex.

Assembling the two doublets into a column vector
\begin{equation}
  \mathcal{H} = \begin{bmatrix} H_1 \\ H_2 \end{bmatrix},
\end{equation}
the kinetic terms and gauge interactions take the form,
\begin{equation}
  \mathcal{L}_{\mathrm{kin}} = (D_\mu \mathcal{H})^\dagger (D^\mu \mathcal{H}) - \tfrac{1}{4}W_{\mu\nu}^i W^{i\,\mu\nu} - \tfrac{1}{4}B_{\mu\nu} B^{\mu\nu}.
\end{equation}
This is invariant under the global $\U(2)_H$ acting as $\mathcal{H} \to U\mathcal{H}$ for $U \in \U(2)_H$. Since $\U(2)_H \cong \SU(2)_H \times \U(1)_H$, and the $\U(1)_H$ factor is identified with hypercharge (both doublets carry $Y = \tfrac{1}{2}$), the family-nontrivial part is the $\SU(2)_H$ subgroup.

\subsection{Invariants}

To organize the potential in $\SU(2)_H$ representations, we introduce the Hermitian $2\times 2$ matrix 
\begin{equation}
  K_{ab} \equiv H_b^\dagger H_a = \frac{S}{2}\,\mathbf{1}_{ab} + \frac{T^i}{2}\,\sigma^i_{ab}, \quad a,b \in \{1,2\},
\label{eq:Higgs_bilinear}
\end{equation}
where $\sigma^i$ are the Pauli matrices.  The trace and the triplet components are~\cite{Ferreira:2010yh,Maniatis:2006fs}
\begin{align}
  S &= \Tr K = |H_1|^2 + |H_2|^2, \\
  T^i &= \Tr(K\sigma^i) =
  \begin{cases}
    H_1^\dagger H_2 + H_2^\dagger H_1 & i=1,\\
    -i(H_1^\dagger H_2 - H_2^\dagger H_1) & i=2,\\
    |H_1|^2 - |H_2|^2 & i=3.
  \end{cases}
\label{eq:Higgs_decomp}
\end{align}
Under $\mathcal{H}\to V\mathcal{H}$ with $V\in\SU(2)_H$, the quantity $S$ is a singlet ($\mathbf{1}$) while $T^i$ rotates as a triplet ($\mathbf{3}$). There is also the antisymmetric $\SU(2)_L$ contraction $S_\epsilon\equiv H_1^T i\sigma_2 H_2$, but $S_\epsilon^\dagger S_\epsilon = \tfrac{1}{2}[(\Tr K)^2 - \Tr K^2]$ is already expressible in terms of $S$ and $T^i$, so it generates no new quartic invariant.

\subsection{Spurions}

The bilinear content of the potential transforms in $\bar{\mathbf{2}}\otimes\mathbf{2} = \mathbf{1}\oplus\mathbf{3}$, giving~\cite{Ferreira:2010yh,Maniatis:2006fs}
\begin{equation}
  V_2 = M_0\, S + M_i\, T^i,
\label{eq:V2}
\end{equation}
with
\begin{align}
    M_0 &= \frac{m_{11}^2+m_{22}^2}{2},
    \nonumber\\
    M_i &= \Bigl(-\Re m_{12}^2,\; \Im m_{12}^2,\; \frac{m_{11}^2-m_{22}^2}{2}\Bigr).
\label{eq:Mcoeffs}
\end{align}
We now make $V_2$ formally invariant under SU(2)$_H$ by promoting $M_0$ and $M_i$ to spurions. Then $M_0$ is the family-singlet quadratic spurion and $M_i$ the family-triplet quadratic spurion.

The quartic potential is built from $(\bar{\mathbf{2}}\otimes\mathbf{2})^{\otimes 2}$,
\begin{equation}
    (\mathbf{1}\oplus\mathbf{3})\otimes(\mathbf{1}\oplus\mathbf{3})
    = \mathbf{1}\oplus\mathbf{3}\oplus\mathbf{3}\oplus(\mathbf{1}\oplus\mathbf{3}\oplus\mathbf{5}).
\label{eq:quartictensor}
\end{equation}
The first two triplets on the right-hand side both arise from the same $T^i$ (one from $\mathbf{1}\otimes\mathbf{3}$ and one from $\mathbf{3}\otimes\mathbf{1}$), so they are not independent. Moreover, because $T^i$ is built from bilinears of commuting bosonic fields, $T^iT^j$ is automatically symmetric, so the antisymmetric triplet in $\mathbf{3}\otimes\mathbf{3}$ is absent. The independent field quartics are therefore
\begin{align}
  S^2 \sim \mathbf{1}, \qquad  &ST^i \sim \mathbf{3}, \qquad  T^2 \equiv \delta_{ij}T^iT^j \sim \mathbf{1}, \qquad \nonumber \\
  &Q^{ij} \equiv T^{(i}T^{j)} - \frac{\delta^{ij}}{3}T^2 \sim \mathbf{5},
\end{align}
where $Q^{ij}$ is symmetric and traceless by construction. The quartic
potential is then
\begin{align}
    \!\!\!\! V_4 = \frac14\Lambda_{00}\,S^2 + \frac12\Lambda_{0i}\,ST^i + \frac{1}{12}\Lambda_{\rm tr}\,T^2 + \frac14\Lambda_{ij}\,Q^{ij} \;.
\label{eq:V4}
\end{align}
We now make $V_4$ formally invariant under SU(2)$_H$ by promoting the quartic couplings to spurions. Then the spurionic couplings $\Lambda_{00}\sim\mathbf{1}$, $\Lambda_{0i}\sim\mathbf{3}$, $\Lambda_{\rm tr}\sim\mathbf{1}$, and $\Lambda_{ij}\sim\mathbf{5}$ (symmetric traceless, $\Lambda_{ii}=0$) carry the same $\SU(2)_H$ quantum numbers as their field-quartic partners. In total, the quartic spurions decompose as
\begin{equation}
    \mathbf{1}\oplus\mathbf{3}\oplus\mathbf{1}\oplus\mathbf{5}.
\end{equation}
Their explicit expressions in terms of the standard 2HDM couplings are
\begin{align}
  \Lambda_{00} &= \frac{\lambda_1+\lambda_2+2\lambda_3}{2},
  \label{eq:Lam00}
  \\
  \Lambda_{0i} &= \Bigl(\Re(\lambda_6+\lambda_7),\, -\Im(\lambda_6+\lambda_7),\, \tfrac{\lambda_1-\lambda_2}{2}\Bigr),
  \label{eq:Lam0i}
  \\
  \Lambda_{\rm tr} &= 2\lambda_4 + \frac{\lambda_1+\lambda_2-2\lambda_3}{2}.
  \label{eq:Lamtr}
\end{align}
The quintet spurion $\Lambda_{ij}$ is obtained from the full symmetric coupling tensor
\begin{align}
  \tilde\Lambda_{ij} =
  \begin{pmatrix}
    \lambda_4+\Re\lambda_5 & -\Im\lambda_5 & \Re(\lambda_6-\lambda_7) \\
    -\Im\lambda_5 & \lambda_4-\Re\lambda_5 & -\Im(\lambda_6-\lambda_7) \\
    \Re(\lambda_6-\lambda_7) & -\Im(\lambda_6-\lambda_7) &
    \dfrac{\lambda_1+\lambda_2-2\lambda_3}{2}
  \end{pmatrix},
\label{eq:Lamtilde}
\end{align}
whose trace is $\tilde\Lambda_{ii} = \Lambda_{\rm tr}$, by removing the trace part,
\begin{align}
  \Lambda_{ij} = \tilde\Lambda_{ij} - \frac{\Lambda_{\rm tr}}{3}\,\delta_{ij},
  \qquad \Lambda_{ii} = 0.
\label{eq:Lamij}
\end{align}

\subsection{Transformation Laws}

A Higgs-family rotation acts on the doublets as $H_a\to U_{ab}H_b$ with $U\in\SU(2)_H$. Every object in the potential is built from bilinears of the doublets, and bilinears see only the adjoint action of $U$. It is therefore convenient to trade $U\in\SU(2)_H$ for the real rotation $R\in \SO(3)$ it induces on the Pauli basis~\cite{Ferreira:2010yh,Maniatis:2006fs},
\begin{align}
    \!\!\!\! U^\dagger\sigma^i U = R_{ij}\,\sigma^j, \qquad R_{ij}=\tfrac12\Tr\!\left(U^\dagger\sigma^i U\,\sigma^j\right),
\label{eq:R_from_U}
\end{align}
with $R^{T}R=\mathbf 1$. This is the familiar two-to-one map $\SU(2)_H\to \SO(3)$.  The Hermitian bilinear transforms in the adjoint, $K\rightarrow U K U^\dagger$, and its singlet and triplet parts inherit their transformations directly. Using the trace form of $S$ and $T^i$ together with Eq.~\eqref{eq:R_from_U},
\begin{align}
    S &\rightarrow S, & T^i &\rightarrow R_{ij}\,T^j.
\end{align}
Every spurion transforms in the same representation as its field-bilinear partner, since the potential is invariant. In the quadratic sector,
\begin{align}
    M_0 &\rightarrow M_0, & M_i &\rightarrow R_{ij}\,M_j,
\end{align}
and in the quartic sector,
\begin{align}
    \Lambda_{00} &\rightarrow \Lambda_{00}, & \Lambda_{\rm tr} &\rightarrow \Lambda_{\rm tr}, \nonumber\\
    \Lambda_{0i} &\rightarrow R_{ij}\,\Lambda_{0j}, & \Lambda_{ij} &\rightarrow R_{ik}R_{jl}\,\Lambda_{kl}.
\end{align}
Note that the triplet index is carried by a real $\SO(3)$ vector, so the metric is simply $\delta_{ij}$.

\subsection{The fixed surface}

The RG-stable relations identified in Refs.~\cite{Ferreira:2023dke,Grzadkowski:2026gkx} are
\begin{equation}
  m_{11}^2+m_{22}^2=0, \qquad \lambda_1=\lambda_2, \qquad \lambda_6=-\lambda_7.
\label{eq:goofysurface}
\end{equation}
In the spurion language developed above, these conditions acquire a simple group-theoretic meaning. Using Eqs.~\eqref{eq:Mcoeffs} and~\eqref{eq:Lam0i}, one can verify that Eq.~\eqref{eq:goofysurface} is equivalent to the vanishing of the quadratic singlet spurion and the quartic triplet spurion,
\begin{equation}
    M_0 = 0, \qquad \Lambda_{0i}=0,
\label{eq:spurionsurface}
\end{equation}
with all other parameters left unrestricted. The scalar potential on the surface~\eqref{eq:spurionsurface} therefore contains a quadratic triplet, two quartic singlets, and one quartic quintet, but no quadratic singlet and no quartic triplet.

\section{Warm-up: a complex scalar toy model} \label{sec:toymodel}

Before addressing the 2HDM, we illustrate the spurion mechanism in the simpler setting of Ref.~\cite{Haber:2025cbb}.  Consider one complex scalar field $\Phi$ with the $\mathbb Z_2$ symmetry $\Phi\to -\Phi$, so that the most general renormalizable potential is
\begin{align}
    V(\Phi)&=m_1^2\,\Phi^\dagger \Phi +\left[\frac{1}{2}m_2^2\,\Phi^2+\hc\right] 
    \\
    &+\lambda_1(\Phi^\dagger\Phi)^2 +\left[\frac{1}{4}\lambda_2\,\Phi^4 + \frac{1}{2}\lambda_3\,\Phi^2\Phi^\dagger\Phi + \hc\right].\nonumber
\label{eq:toycomplex}
\end{align}
All parameters may be complex, except $m_1^2$ and $\lambda_1$, which are real by hermiticity. As shown in Ref.~\cite{Haber:2025cbb}, the conditions
\begin{equation}
    m_1^2=0, \qquad \lambda_3=0,
\label{eq:U1surface}
\end{equation}
define an RG-invariant surface. For general $m_2^2$ and $\lambda_2$, however, no actual symmetry of the full potential enforces both conditions simultaneously.

The explanation in terms of spurions is straightforward. The kinetic term is invariant under a global $\U(1)$ transformation, $\Phi \to e^{i\alpha}\Phi$, and we treat this $\U(1)$ as a spurionic symmetry of the full theory. Each coupling is thus assigned a spurionic charge:
\begin{align}
    q(\Phi)&=+1, &  q(m_1^2)&=0, & q(m_2^2)&=-2, \nonumber\\
    q(\lambda_1)&=0, & q(\lambda_2)&=-4, & q(\lambda_3)&=-2.
\end{align}

In a mass-independent renormalization scheme, the beta functions are polynomial in the couplings and must respect the spurionic $\U(1)$ charge. Moreover, dimensional analysis implies that $\beta_{m_1^2}$ is linear in the mass spurions. Since $\beta_{m_1^2}$ has charge zero, its general form is
\begin{align}
  \beta_{m_1^2} =  m_1^2\,F_0 +m_2^2\left(\lambda_3^*F_1+\lambda_2^*\lambda_3F_2\right) + \text{c.c.}
\end{align}
where $F_{0,1,2}$ are arbitrary charge-neutral polynomial functions of the dimensionless spurions, built from combinations such as
\begin{align}
    \lambda_1,\quad  |\lambda_2|^2,\quad  |\lambda_3|^2,\quad \lambda_2(\lambda_3^*)^2,\quad \lambda_2^*\lambda_3^2,\quad \ldots
\end{align}
Therefore, every contribution to $\beta_{m_1^2}$ vanishes when $m_1^2=\lambda_3=0$.

Similarly, $\beta_{\lambda_3}$ must carry charge $-2$. The available couplings have charges $q(\lambda_1)=0$, $q(\lambda_2)=-4$, $q(\lambda_3)=-2$, $q(\lambda_3^*)=+2$. Any charge-$(-2)$ monomial built from these must contain at least one factor of $\lambda_3$, so the most general form is
\begin{equation}
    \beta_{\lambda_3}= \lambda_3 G_1+\lambda_2\lambda_3^*G_2,
\end{equation}
where $G_{1,2}$ are charge-neutral polynomial functions of the dimensionless spurions. Therefore, every term vanishes when $\lambda_3=0$, and
\begin{equation}
  \beta_{\lambda_3}\big|_{\lambda_3=0}=0.
\end{equation}
Charge conservation holds at every order in perturbation theory, so the same counting applies to all loops. Therefore, the surface~\eqref{eq:U1surface} is RG-invariant to all orders.

As we shall see, this is the direct analogue of the closure condition in the 2HDM. In the toy model, the crucial fact is the absence of any spurion with charge $-2$ once $\lambda_3$ is set to zero. In the 2HDM, the same mechanism appears in a non-Abelian guise, where the role of the missing charge $-2$ spurion is played by the missing family-triplet spurion.

\section{Leading-order proof in the 2HDM} \label{sec:firstorder}

We first show that, in the pure scalar sector, the surface
\begin{equation}
  \mathcal{S}:\quad M_0 = 0,\quad \Lambda_{0i} = 0
\label{eq:surfS}
\end{equation}
is one-loop RG-invariant. This follows directly from elementary $\SU(2)_H$ representation theory.

\subsection{Covariant structure of the beta functions}

At one loop, it is useful to separate the purely scalar contributions from those involving gauge fields. For a purely scalar contribution to a quartic beta function, the one-loop diagram contains two quartic vertices. Indeed, with $V$ quartic vertices, $P$ internal scalar propagators, and $E=4$ external scalar legs,
\begin{align}
L=P-V+1, \qquad 4V=2P+E ,
\end{align}
so $L=1$ implies $V=2$. The purely scalar part of a one-loop quartic beta function is therefore quadratic in the quartic spurions. Similarly, the one-loop purely scalar contribution to a quadratic beta function is linear in a quadratic spurion and linear in a quartic spurion. 

The beta function of any spurion must transform in the same representation as the spurion itself. Consider first $\beta_{\Lambda_{0i}}$, which must be a triplet. By dimensional analysis, no mass spurions can appear in a quartic beta function at any loop order in a mass-independent scheme. The available quartic spurions are therefore two singlets $\Lambda_{00}$ and $\Lambda_{\rm tr}$, one triplet $\Lambda_{0i}$, and one quintet $\Lambda_{ij}$.

It is convenient to first enumerate the products of two quartic spurions without imposing $\Lambda_{0i}=0$. Products of two singlets give only singlets, while a singlet times a quintet gives only a quintet and neither can contribute to the beta function of the triplet. The mixed products $\mathbf1\otimes\mathbf3=\mathbf3$ and $\mathbf3\otimes\mathbf5\supset\mathbf3$ do contain triplets, giving structures proportional to $\Lambda_{00}\Lambda_{0i}$, $\Lambda_{\rm tr}\Lambda_{0i}$, and $\Lambda_{ij}\Lambda_0^j$. It remains to examine products of two triplet spurions and products of two quintet spurions. Since the components of each spurion commute, these products lie in $\Sym^2(\mathbf3)$ and $\Sym^2(\mathbf5)$, respectively. We show below that neither contains a triplet.

\paragraph{$\Sym^2(\mathbf{3})$ contains no triplet.}
A triplet in $\mathbf{3}\otimes\mathbf{3}$ would require the antisymmetric combination $\epsilon^{ijk}\Lambda_{0j}\Lambda_{0k}$, which however vanishes identically. The symmetric product therefore decomposes as
\begin{equation}
  \Sym^2(\mathbf{3}) = \mathbf{1}\oplus\mathbf{5},
\end{equation}
with no triplet.

\paragraph{$\Sym^2(\mathbf{5})$ contains no triplet.}
Any triplet built from two quintets $\Lambda_{ij}$ must use the invariant $\epsilon^{ijk}$ to carry the free triplet index. The unique candidate contraction is $\epsilon^{ijk}\Lambda_{jl}\Lambda_{k}^{l}$. However, $\Lambda_{jl}\Lambda_{k}^{l}$ is symmetric under $j\leftrightarrow k$ while $\epsilon^{ijk}$ is antisymmetric, so the contraction therefore vanishes identically. Hence,
\begin{equation}
  \Sym^2(\mathbf{5}) = \mathbf{1}\oplus\mathbf{5}\oplus\mathbf{9},
\label{eq:sym2_5}
\end{equation}
with no triplet. Since the gauge couplings carry no Higgs-family indices and are therefore $\SU(2)_H$ singlets, they do not affect the representation-theoretic argument above.

It follows that on the surface $\mathcal{S}$, every term in $\beta_{\Lambda_{0i}}$ is proportional to $\Lambda_{0i}$. The most general covariant form at one loop is therefore
\begin{equation}
  \beta_{\Lambda_{0i}} = b_1\,\Lambda_{00}\,\Lambda_{0i} + b_2\,\Lambda_{\mathrm{tr}}\,\Lambda_{0i} + b_3\,\Lambda_{ij}\,\Lambda_0^j + \cdots,
\label{eq:betaLambda}
\end{equation}
where the ellipsis denotes singlet dressings from gauge couplings, which do not affect the representation-theoretic argument. Hence
\begin{equation}
  \Lambda_{0i}=0 \;\Longrightarrow\; \beta_{\Lambda_{0i}}=0.
\end{equation}

Turn now to $\beta_{M_0}$, which must be an $\SU(2)_H$ singlet of mass dimension two. By dimensional analysis, exactly one mass spurion ($M_0$ or $M_i$) and one quartic spurion must appear. The most general covariant form is given by, 
\begin{equation}
  \beta_{M_0} = a_1\,M_0\,\Lambda_{00} + a_2\,M_0\,\Lambda_{\mathrm{tr}} + a_3\,M_i\,\Lambda_0^i + \cdots.
\label{eq:betaM0}
\end{equation}
The term $M_i\Lambda_0^i$ is the unique singlet arising from the contraction of the quadratic and quartic triplets. On the surface $\mathcal{S}$ it vanishes with $\Lambda_{0i}$, and the remaining terms vanish with $M_0$, so
\begin{equation}
  M_0=\Lambda_{0i}=0 \;\Longrightarrow\; \beta_{M_0}=0.
\end{equation}

The surface $\mathcal S$ is therefore closed under the scalar RG flow at lowest order. Gauge couplings are $\SU(2)_H$ singlets and do not alter this conclusion. Yukawa interactions introduce additional non-singlet $\SU(2)_H$ structures and require a separate analysis, carried out in Sec.~\ref{sec:yukawa}.\\

\subsection{Explicit one-loop verification}

At one-loop, and suppressing fermionic contributions for the moment, the combinations of beta functions derived in Ref.~\cite{Ferreira:2023dke} are
\begin{widetext}
\begin{align}
      \beta_{m_{11}^2+m_{22}^2} ={}& 3(\lambda_1 m_{11}^2 + \lambda_2 m_{22}^2) +\left(2\lambda_3+\lambda_4 -\frac94 g^2-\frac34g'^2\right)(m_{11}^2+m_{22}^2) -3\bigl[(\lambda_6^*+\lambda_7^*)m_{12}^2+\hc\bigr],
      \label{eq:betaM0raw}
      \\[4pt]
      \beta_{\lambda_1-\lambda_2} ={}&  6(\lambda_1^2-\lambda_2^2) +12\bigl(|\lambda_6|^2-|\lambda_7|^2\bigr) -\frac{3}{2}(3g^2+g'^2)(\lambda_1-\lambda_2),
      \label{eq:betal1l2raw}
      \\[4pt]
      \beta_{\lambda_6+\lambda_7} ={}& 6(\lambda_1\lambda_6+\lambda_2\lambda_7) +6(\lambda_3+\lambda_4)(\lambda_6+\lambda_7) +6\lambda_5(\lambda_6^*+\lambda_7^*) -\frac{3}{2}(3g^2+g'^2)(\lambda_6+\lambda_7).
      \label{eq:betal6l7raw}
\end{align}
\end{widetext}

These equations make the one-loop fixed surface completely explicit. If $\lambda_1=\lambda_2$ and $\lambda_6=-\lambda_7$, then
\begin{equation}
  \beta_{\lambda_1-\lambda_2}=0, \qquad \beta_{\lambda_6+\lambda_7}=0.
\end{equation}
Once these are imposed, Eq.~\eqref{eq:betaM0raw} reduces to
\begin{equation}
    \beta_{m_{11}^2+m_{22}^2} = \Big[3\lambda_1 + 2\lambda_3+\lambda_4 - \frac{1}{4}(9g^2+3g'^2)\Big](m_{11}^2+m_{22}^2),
\end{equation}
so that $m_{11}^2+m_{22}^2=0$ is preserved as well. This is the explicit one-loop check of Ref.~\cite{Ferreira:2023dke}.

The one-loop formulas above are simply the component form of this covariant statement. The remaining question is whether the same conclusion holds at higher orders, to which we now turn.

\section{All-orders closure in the scalar sector} \label{sec:allorders}

We now extend the one-loop argument to all orders in the purely scalar sector. On the surface
\begin{equation}
    \mathcal{S}:\qquad M_0=0,\qquad \Lambda_{0i}=0,
\end{equation}
the remaining quartic spurions are the two singlets $\Lambda_{00}$, $\Lambda_{\rm tr}$ and the lone quintet $\Lambda_{ij}$, while the remaining quadratic spurion is the triplet $M_i$. We analyze the beta functions of $\Lambda_{0i}$ and $M_0$ in turn and show that, in both cases, their vanishing reduces to the same group-theoretic condition.

\paragraph{Beta function of $\Lambda_{0i}$.}

In a mass-independent scheme, $\beta_{\Lambda_{0i}}$ can depend only on dimensionless couplings. Gauge couplings carry no Higgs-family indices and therefore enter only as $\SU(2)_H$ singlet dressings. The same is true of the Higgs-family singlet quartic spurions $\Lambda_{00}$ and $\Lambda_{\rm tr}$. Thus a general contribution independent of $\Lambda_{0i}$ transforms schematically as
\begin{align}
    \Sym^{a}(\mathbf 1)\otimes \Sym^{c}(\mathbf 5),
\end{align}
where the $\mathbf 1$ collectively denotes all Higgs-family singlet factors, including the gauge couplings. Since $\Sym^{a}(\mathbf 1)=\mathbf 1$, these factors do not affect the nontrivial representation content. Consequently, the question of whether such a term can source $\beta_{\Lambda_{0i}}$ reduces entirely to whether
\begin{align}
    \mathbf 3 \subset \Sym^{c}(\mathbf 5)
\end{align}
for any $c\geq0$. The most general beta-function for the triplet can be written as,
\begin{align}
    \beta_{\Lambda_{0i}} = \Lambda_{0j}\,F_{ij} + G_i,
\label{eq:betaLam0_split}
\end{align}
where $F_{ij}$ and $G_i$ are covariants built from the singlets and the quintet. The first term collects every contribution proportional to $\Lambda_{0i}$ and vanishes on $\mathcal S$. The second term $G_i$ carries no factor of $\Lambda_{0i}$ and is the only structure that \textit{might} survive. For this, we would need to check whether
\begin{align}
    \mathbf 3 \subset \Sym^{c}(\mathbf 5).
\label{eq:keyclaim}
\end{align}

\paragraph{Beta function of $M_0$.} 

In a mass-independent scheme, $\beta_{M_0}$ is a Higgs-family singlet of mass dimension two and is linear in the quadratic spurions. Since $M_0\sim\mathbf{1}$ and $M_i\sim\mathbf{3}$, its general structure is
\begin{equation}
    \beta_{M_0} = M_0\,F_{\mathbf 1} + M_i\,G^{i},
\label{eq:betaM0all}
\end{equation}
where $F_{\mathbf 1}$ is a singlet and $G^i$ a triplet function of the dimensionless scalar and gauge couplings. The term proportional to $M_0$ vanishes on $\mathcal{S}$. The term proportional to $M_i$ only contributes if $\Sym^c(\mathbf 5)$ contains a $\mathbf{3}$. This follows from the $\SU(2)$ identity $\mathbf{3}\otimes R\supset\mathbf{1}$ if and only if  $R\supset\mathbf{3}$, which is an immediate consequence of  character orthonormality. Hence $G^i$ can be nonzero on $\mathcal{S}$ only if $\Sym^{c}(\mathbf{5})$ contains a $\mathbf{3}$, which is again precisely the condition~\eqref{eq:keyclaim}.

The vanishing of both beta functions, Eqs.~\eqref{eq:betaLam0_split} and \eqref{eq:betaM0all}, reduces to the same condition that the all-orders closure of $\mathcal{S}$ in the scalar sector holds if and only if
\begin{equation}
    \mathbf{3}\not\subset \Sym^{n}(\mathbf{5}) \qquad \text{for all } n\ge 0.
\end{equation}
We prove this in Appendix~\ref{app:characters} using a character-theory generating functional. 

Granting this result, $G^i=0$ whenever $\Lambda_{0i}=0$. We conclude that once $M_0$ and $\Lambda_{0i}$ are set to zero, no scalar loop correction can regenerate them, establishing the desired result.

\section{Yukawa spurion decomposition} \label{sec:yukawa}

The all-orders argument of Sec.~\ref{sec:allorders} was established within the scalar-gauge sector. Yukawa interactions enlarge the set of Higgs-family covariants available to the flow and therefore reopen the closure problem, since fermion bilinears can source the covariant directions whose vanishing defines $\mathcal{S}$. Fermionic extensions of the GOOFy construction have recently been studied in Refs.~\cite{Grzadkowski:2026gkx,Ferreira:2026ccg}. Here we begin by casting the Yukawa sector in a form that makes its Higgs-family transformation properties manifest. The Yukawa terms in the Lagrangian take the form
\begin{align}
    \!\!\mathcal L_Y &= \bar Q_L\big(\Gamma_1 H_1+\Gamma_2 H_2\big)d_R + \bar Q_L\big(\Delta_1 \widetilde H_1+\Delta_2 \widetilde H_2\big)u_R \nonumber
    \\
    &\quad + \bar L_L\big(\Pi_1 H_1+\Pi_2 H_2\big)e_R + \mathrm{h.c.},
\label{eq:LY}
\end{align}
where $\Gamma_a$, $\Delta_a$, and $\Pi_a$ are matrices in fermion-generation space~\cite{Davidson:2005cw,Haber:2006ue}, and $\widetilde H_a = i\sigma_2 H_a^*$ is the conjugate doublet. Under a Higgs-family transformation, $H_a\rightarrow U_{ab}H_b$, $\widetilde H_a\rightarrow U_{ab}^{*}\widetilde H_b$ with $U\in \SU(2)_H$. Then, invariance of Eq.~\eqref{eq:LY} under the spurionic $\SU(2)_H$ symmetry requires $\Gamma_a\rightarrow U_{ab}^{*}\Gamma_b$, $\Pi_a\rightarrow U_{ab}^{*}\Pi_b$, $\Delta_a\rightarrow U_{ab}\Delta_b$. Since the fundamental representation of $\SU(2)$ is pseudo-real, the first two may be converted into ordinary doublets. Defining
\begin{equation}
    \mathcal Y_a^{(u)}\equiv\Delta_a,
    \quad
    \mathcal Y_a^{(d)}\equiv\epsilon_{ab}\Gamma_b,
    \quad
    \mathcal Y_a^{(e)}\equiv\epsilon_{ab}\Pi_b,
\label{eq:common_yukawa_doublets}
\end{equation}
with $\epsilon_{12}=1$, gives $\mathcal Y_a^{(f)} \rightarrow U_{ab}\mathcal Y_b^{(f)}$ for every fermion species $f\in \{u,d,e\}$. Eq.~\eqref{eq:common_yukawa_doublets} therefore places all Yukawa spurions in a common Higgs-family convention.

In the Yukawa sector, we again build a Higgs-family bilinear in direct analogy with Eq.~\eqref{eq:Higgs_bilinear}. There the two doublets were contracted into a number in each Higgs-family entry, whereas here we contract the two Yukawa fundamentals while leaving their generation indices open,
\begin{align}
    \mathsf W_{ab}^{(f)}  \equiv \mathcal Y_a^{(f)}\mathcal Y_b^{(f)\dagger},
    \qquad
    \mathsf W_{ab}^{(f)\dagger} = \mathsf W_{ba}^{(f)},
\label{eq:yukawa_bilinear}
\end{align}
so that each entry $\mathsf W_{ab}^{(f)}$ is itself a matrix acting on the left-handed generations. The object $\mathsf W^{(f)}$ carries two Higgs-family indices and transforms by conjugation, $\mathsf W^{(f)}\to U\mathsf W^{(f)}U^\dagger$, following from $\mathcal Y_a^{(f)}\to U_{ab}\mathcal Y_b^{(f)}$. It therefore reduces under $\SU(2)_H$ into a singlet and a triplet,
\begin{align}
    \mathsf W_{ab}^{(f)} = \frac12 \left( \delta_{ab}\mathsf W_0^{(f)} + \sigma^i_{ab}\mathsf W_i^{(f)} \right),
\label{eq:W_decomposition}
\end{align}
with $\mathsf W_0^{(f)}\sim\mathbf 1$ and $\mathsf W_i^{(f)}\sim\mathbf 3$. The distinction from the scalar case is that the coefficients $\mathsf W_0^{(f)}$ and $\mathsf W_i^{(f)}$ are Hermitian matrices in generation space rather than numbers. 

Since we are computing the beta functions of scalar masses and couplings, the fermions enter only through closed loops~\cite{Machacek:1984zw,Ferreira:2010xe}, and the Higgs-family content of any such contribution is a product of the Yukawa bilinears $\mathsf W^{(f)}_{ab}$. A single Yukawa bilinear already carries a triplet $\mathsf W^{(f)}_i$, and products of Yukawa bilinears appearing at higher loop order generate higher-spin covariants as well. The covariants whose absence protected the scalar-gauge surface are thus regenerated by the Yukawa couplings. These triplets can contribute directly to $\beta_{\Lambda_{0i}}$, while the higher-spin covariants can combine among themselves or with $\Lambda_{ij}$ to produce a triplet. Consequently the GOOFy surface is not closed for general Yukawa couplings.

We therefore do not seek texture-by-texture cancellations. Instead, let $G$ be an exact symmetry of the full 2HDM Lagrangian.\footnote{Our concern here is purely group-theoretic. We ask whether imposing $G$ guarantees RG closure of the surface, without requiring that the corresponding $G$-symmetric Yukawa sector be phenomenologically viable.} 
We then ask whether the surface remains closed once restricted to the $G$-invariant locus. This is the question of whether the flow preserves
\begin{align}
    \mathcal S \cap G:\qquad M_0=0,\qquad \Lambda_{0i}=0.
\label{eq:surfSG}
\end{align}
The argument follows directly once $G$ is imposed on the full theory. An exact non-anomalous symmetry is respected by the renormalized effective action, and therefore by its counterterms and beta functions.
A spurionic coupling transforming in a Higgs-family representation that contains no $G$-invariant operator cannot be generated at any order in perturbation theory.

Both surface conditions follow from this. The spurionic coupling $\Lambda_{0i}$ and its beta function $\beta_{\Lambda_{0i}}$ are real Higgs-family triplets. If $G$ admits no invariant real triplet operator, then $\Lambda_{0i}$ cannot be generated at any order, and $\beta_{\Lambda_{0i}}$ vanishes identically on the $G$-symmetric locus. The same reasoning protects $M_0$. Because $\beta_{M_0}$ is linear in the mass parameters, the singlet $M_0$ can be fed from the triplet $M_i$ only by contracting $M_i$ against a triplet coupling. With no invariant triplet available, that contraction cannot be formed, and $M_0=0$ is preserved. The detailed Yukawa tensors do not alter this conclusion. Because they themselves respect $G$, every combination of them entering a scalar counterterm obeys the same selection rule.

\subsection{The six symmetry classes}\label{sec:six_classes}

The 2HDM scalar sector admits six symmetry classes that can be imposed on the Higgs doublets~\cite{Ivanov:2005hg,Ivanov:2006yq,Peccei:1977hh}: three Higgs-family symmetries ($\mathbb Z_2$, $\U(1)_{PQ}$, and $\SO(3)_H$) and three Generalized-$\CP$ (GCP) symmetries ($\CP1$, $\CP2$, and $\CP3$)~\cite{Ecker:1987qp,Neufeld:1987wa,Ferreira:2010bm}. For each symmetry class, we find the image $R(G)$ of $G$ acting on the Higgs-family triplet and determine whether it leaves any direction fixed. Where no direction survives, the triplet contains no $G$-invariant operator, and $\Lambda_{0i}$ together with every Yukawa covariant along it is forbidden to all orders. Where a direction survives, $G$ does not immediately forbid all components of the triplet operator, and we must determine separately whether the Yukawa sector generates a covariant along that direction. 

\subsubsection*{Generalized-$\CP$ Classes}
We treat the GCP classes together, since they differ only in the unitary matrix that dresses the conjugation. A general GCP transformation combines a spacetime parity reflection $x=(t,\vec x)\to x_P=(t,-\vec x)$ with a complex conjugation and a Higgs-family rotation,
\begin{align}
    H_a(x) \rightarrow X_{ab}\,H_b^*(x_P), \qquad X\in U(2).
\label{eq:GCP_def}
\end{align}
Since the parity reflection acts only on the spacetime argument and plays no role in the Higgs-family analysis, we suppress it from here on and track only the action on the family indices, $H_a\to X_{ab}H_b^*$. The bilinear then transforms as
\begin{align}
    K\rightarrow(XKX^\dagger)^*.
\label{eq:GCP_K}
\end{align}
Conjugation by $X$ leaves the singlet unchanged and rotates the Pauli matrices according to $X\sigma^iX^\dagger=R_{ji}(X)\sigma^j$ with $R_X\in \SO(3)$. Complex conjugation then leaves $\sigma^1$ and $\sigma^3$ unchanged while reversing the sign of $\sigma^2$, that is $(\sigma^i)^*=(\Pi_2)_{ji}\sigma^j$ with $\Pi_2=\mathrm{diag}(1,-1,1)$. In terms of the Higgs-bilinears $S$ and $T^i$ introduced in Eqs.~\eqref{eq:Higgs_bilinear}-\eqref{eq:Higgs_decomp},
\begin{align}
    S \rightarrow S, \qquad T^i\rightarrow R(G)_{ij}T^j,
\label{eq:GCP_T}
\end{align}
where $R(G)=\Pi_2 R_X$. The action of a generalized $\CP$ transformation on the triplet is thus an ordinary rotation induced by $X$ followed by the reflection associated with complex conjugation.

\paragraph{$\CP1$.} Here the unitary matrix is trivial, $X=\mathbf 1$, so $R_X=\mathbf 1$ and
\begin{align}
    R(\CP1)=\Pi_2=\mathrm{diag}(1,-1,1).
\end{align}
This leaves the $T^1$ and $T^3$ directions fixed and reverses $T^2$.The surviving invariant subspace is therefore the plane orthogonal to $\hat 2$, so $\CP1$ forces only the $\CP$-odd component $\Lambda_{02}$ to vanish, while $\Lambda_{01}$ and $\Lambda_{03}$ remain allowed. The Yukawa sector can populate both of these surviving directions. At lowest order, the corresponding Yukawa structures are
\begin{align}
  &2\,\mathrm{Re}\,\Tr_F\!\left(\mathcal Y_1^{(f)}\mathcal Y_2^{(f)\dagger}\right),
  \\
  &\Tr_F\!\left(\mathcal Y_1^{(f)}\mathcal Y_1^{(f)\dagger}\right) - \Tr_F\!\left( \mathcal Y_2^{(f)}\mathcal Y_2^{(f)\dagger} \right),
\label{eq:CP1_covariants}
\end{align}
which transform along the $T^1$ and $T^3$ directions, respectively. The corresponding $T^2$ structure,
\begin{align}
  2\,\mathrm{Im}\,\Tr_F\!\left(\mathcal Y_1^{(f)}\mathcal Y_2^{(f)\dagger} \right),
\end{align}
is $\CP1$ odd and hence vanishes on the $\CP1$-symmetric locus. The first of the allowed structures is expected to be nonzero when both doublets couple to the same fermion species, while the second is also generally nonvanishing. They therefore provide symmetry-allowed sources for the surviving triplet components in the scalar beta functions. Thus $\CP1$ protects only the $\CP$-odd component and not the full GOOFy relation $\Lambda_{0i}=0$. Therefore, the surface does not close.

\paragraph{$\CP2$.} Here $X=i\sigma_2$, whose action on the Pauli matrices is a
rotation by $\pi$ about the second axis, $R_X=\mathrm{diag}(-1,1,-1)$. Hence
\begin{align}
R(\CP2)=\Pi_2 R_X=-\mathbf 1,
\end{align}
and every component of the triplet changes sign. No nonzero $\CP2$-invariant triplet exists, so $\CP2$ forbids the entire triplet sector, requiring both $\Lambda_{0i}=0$ and $M_i=0$. In the Yukawa sector, the generalized-$\CP$ transformation may be accompanied by unitary transformations $V_L^{(f)}$ and $V_R^{(f)}$ acting in generation space on the left- and right-handed fermions of species $f$, respectively. For $\CP2$, invariance of the Yukawa interactions then requires
\begin{align}
    \mathcal Y_1^{(f)} &= V_L^{(f)\dagger}\mathcal Y_2^{(f)}V_R^{(f)}, \nonumber
    \\
    \mathcal Y_2^{(f)} &= -\,V_L^{(f)\dagger}\mathcal Y_1^{(f)}V_R^{(f)} .
\end{align}
Although the corresponding generation-space triplets $\mathsf W_i^{(f)}$ need not vanish individually, any flavor-singlet Yukawa structure entering a scalar counterterm must respect the same $\CP2$ transformation. Since no invariant real Higgs-family triplet exists, such structures cannot source $\Lambda_{0i}$ at any order. Together with the absence of the scalar mass triplet $M_i$, this also preserves $M_0=0$, and the surface is therefore closed to all orders.

\paragraph{$\CP3$.} Here $X$ is a real rotation by an angle $\theta$ in the $(H_1,H_2)$ plane, whose action on the Pauli matrices is a rotation by $2\theta$ about the second axis, $R_X=R_2(2\theta)$. Hence
\begin{align}
    R(\CP3)=\Pi_2 R_2(2\theta),
\end{align}
which reverses $T^2$ while rotating the $(T^1,T^3)$ plane by $2\theta$. For a $\CP3$ transformation, $0<\theta<\pi/2$, no nonzero triplet is left invariant. As with $\CP2$, the triplet sector is forbidden, $\Lambda_{0i}=0$ and $M_i=0$. As in the $\CP2$ case, the generation-space triplets $\mathsf W_i^{(f)}$ need not vanish individually. However, any flavor-singlet Yukawa structure entering a scalar counterterm must respect the same $\CP3$ symmetry. Since no invariant real Higgs-family triplet exists, the Yukawa sector cannot source $\Lambda_{0i}$ at any order. Together with $M_i=0$, this also preserves $M_0=0$, and the surface is therefore closed to all orders.

\subsubsection*{Higgs-family Classes}

The Higgs-family classes act by a unitary transformation alone, $H_a\rightarrow X_{ab}H_b$ with $X\in U(2)$. The induced action on the Higgs-bilinear is ordinary conjugation, $K \rightarrow XK X^\dagger$, which leaves the singlet unchanged and rotates the triplet by the $\SO(3)$ image of $X$,
\begin{align}
    S\rightarrow S,\qquad T^i\rightarrow R(G)_{ij}T^j,
\label{eq:HF_T}
\end{align}
with $R(G)=R_X$. In contrast to the generalized $\CP$ case there is no complex conjugation and hence no reflection $\Pi_2$.

\paragraph{$\SO(3)_H$.} The Higgs-family group is realized in full, so $X$ ranges over all of $\SU(2)$ and its image over the entire rotation group, $R(\SO(3)_H)=\SO(3)$. No nonzero vector is fixed under every rotation, so the triplet sector is forbidden, $\Lambda_{0i}=0$ and $M_i=0$. Exact $\SO(3)_H$ invariance also removes the scalar quintet $\Lambda_{ij}$, leaving only the quartic singlets, so $\mathcal S\cap\SO(3)_H$ is trivially closed to all orders.

\paragraph{$\mathbb Z_2$.} The standard $\mathbb Z_2$ acts as $H_1\to H_1$ and $H_2\to-H_2$. Up to an irrelevant overall phase, this may be represented by $X=i\sigma_3$, whose action on the triplet is
\begin{align}
    R(\mathbb Z_2)=\mathrm{diag}(-1,-1,1),
\end{align}
corresponding to a rotation by $\pi$ about the third axis. Thus $T^1$ and $T^2$ change sign, while $T^3$ is invariant. In the triplet sector, $\mathbb Z_2$ therefore forbids $\Lambda_{01}$ and $\Lambda_{02}$ but leaves $\Lambda_{03}\propto\lambda_1-\lambda_2$ unconstrained. The symmetry also removes the corresponding $\mathbb Z_2$-odd components of the scalar quintet. Together, these conditions give the familiar $\lambda_6=\lambda_7=0$. On the GOOFy surface, however, $\Lambda_{03}=0$ additionally requires $\lambda_1=\lambda_2$, which is not enforced by $\mathbb Z_2$.

The Yukawa sector can populate precisely this surviving triplet direction. Already at the lowest orders, the $\mathbb Z_2$-invariant structures
\begin{align}
    &\Tr_F\!\left(\mathcal Y_1^{(f)}\mathcal Y_1^{(f)\dagger} \right) - \Tr_F\!\left(\mathcal Y_2^{(f)}\mathcal Y_2^{(f)\dagger}\right), 
    \nonumber\\
    &\Tr_F\!\left[\left(\mathcal Y_1^{(f)}\mathcal Y_1^{(f)\dagger}\right)^2 \right] - \Tr_F\!\left[ \left(\mathcal Y_2^{(f)}\mathcal Y_2^{(f)\dagger} \right)^2 \right]
\label{eq:Z2_covariants}
\end{align}
transform along the invariant $T^3$ direction and are generally nonzero for general Yukawa couplings compatible with $\mathbb Z_2$. The first structure provides the lowest-order Yukawa triplet along the surviving $T^3$ direction and enters the quartic beta functions multiplied by scalar quartics, while the second gives a direct $Y^4$ contribution from the one-loop fermion box. For example, in a Type-II assignment the up-type contribution is top dominated,
\begin{align}
    \Tr_F\!\left( \mathcal Y_1^{(u)}\mathcal Y_1^{(u)\dagger}  \right) - \Tr_F\!\left( \mathcal Y_2^{(u)}\mathcal Y_2^{(u)\dagger} \right) &\simeq -y_t^2,
    \nonumber\\
    \Tr_F\!\left[ \left( \mathcal Y_1^{(u)}\mathcal Y_1^{(u)\dagger} \right)^2 \right] - \Tr_F\!\left[ \left( \mathcal Y_2^{(u)}\mathcal Y_2^{(u)\dagger} \right)^2 \right] &\simeq -y_t^4 .
\end{align}
Correspondingly, the one-loop RGE contains $\beta_{\lambda_2}\supset -12\,y_t^4$, with no analogous top contribution to $\beta_{\lambda_1}$, so $\beta_{\Lambda_{03}}\propto\beta_{\lambda_1-\lambda_2}$ does not vanish. The $\mathbb Z_2$ symmetry also allows a mass triplet along the same direction, $M_3\neq0$, and the allowed dimensionless triplet structures can contract with $M_3$ in $\beta_{M_0}$. Hence $M_0=0$ is not protected either, and the GOOFy surface does not close.

\paragraph{$\U(1)_{PQ}$.} The Peccei-Quinn $\U(1)$ acts as $H_1\to e^{i\alpha}H_1$, $H_2\to e^{-i\alpha}H_2$, generated by $X=\mathrm{diag}(e^{i\alpha},e^{-i\alpha})$. Its image is a rotation by $2\alpha$ about the third axis, 
\begin{align}
    R(\U(1)_{PQ})=R_{\hat 3}(2\alpha) \;,
\end{align}
so the $(T^1,T^2)$ plane rotates continuously while $T^3$ is fixed. Demanding invariance under all $\alpha$ leaves the same surviving axis $\hat 3$ as $\mathbb Z_2$. In the quartic quintet, however, the continuous $\U(1)$ imposes the additional condition $\lambda_5=0$, leaving only the single quintet component invariant under rotations about $\hat 3$. 

The Yukawa sector does not remove the surviving triplet direction. For a natural-flavor-conserving Yukawa assignment compatible with $\U(1)_{PQ}$, the structures in Eq.~\eqref{eq:Z2_covariants} remain invariant and can again populate the $T^3$ direction. In particular, for the Type-II assignment the same top-dominated $y_t^4$ contribution to $\beta_{\lambda_1-\lambda_2}$ survives. Thus, although enlarging $\mathbb Z_2$ to $\U(1)_{PQ}$ constrains the quartic sector, it does not eliminate the triplet direction responsible for the failure of closure, and so $\U(1)_{PQ}$ alone does not close the GOOFy surface.

\begin{table}[t]
\centering
\begin{tabular}{lccc}
\toprule
$G$ & action on $T^i$ & invariant direction & closes $\mathcal{S}$ \\
\midrule
$\mathbb Z_2$   & $R_{\hat 3}(\pi)$            & $\hat 3$            & no \\
$\U(1)_{PQ}$     & $R_{\hat 3}(2\alpha)$        & $\hat 3$            & no \\
$\SO(3)_H$      & $\SO(3)$                     & none               & yes \\
$\CP1$          & $\Pi_{\hat 2}$               & $\hat 2^{\perp}$    & no \\
$\CP2$          & $-\mathbf 1$                 & none               & yes \\
$\CP3$          & $\Pi_{\hat 2}R_{\hat 2}(2\theta)$ & none          & yes \\
\bottomrule
\end{tabular}
\caption{Action on the Higgs-family triplet for the six 2HDM symmetry classes. Here $R_{\hat n}(\varphi)$ is a proper rotation about $\hat n$ and $\Pi_{\hat n}$ the reflection reversing $\hat n$. For $\CP3$, $0<\theta<\pi/2$. The surviving directions of $\mathbb Z_2$, $\U(1)_{PQ}$, and $\CP1$ are populated by nonzero Yukawa covariants and the surface fails to close, whereas $\SO(3)_H$, $\CP2$, and $\CP3$ leave no invariant triplet and close $\mathcal S$ to all orders.}
\label{tab:classification}
\end{table}

\subsubsection*{Summary}

Thus the six classes fall into two groups. The symmetries $\SO(3)_H$, $\CP2$, and $\CP3$ leave no nonzero triplet invariant, so they forbid $\Lambda_{0i}$ and $M_i$ and the induced Yukawa covariants alike, closing $\mathcal S$ to all orders. The symmetries $\mathbb Z_2$, $\U(1)_{PQ}$, and $\CP1$ retain a surviving direction that the symmetry does not constrain, and along it the Yukawa sector supplies explicit invariant covariants that regenerate the triplet at one loop. The induced rotations $R(G)$, the surviving directions, and the resulting status of the surface are collected in Table~\ref{tab:classification}.

\section{Conclusions}\label{sec:conclusions}

In this work, we have shown that the RG-stable surface identified in recent 2HDM analyses admits a simple and conventional interpretation in terms of spurions. The gauge-covariant kinetic terms possess a Higgs-family $\U(2)_H$ symmetry, and the scalar couplings organize into irreducible spurions of its $\SU(2)_H$ subgroup. The fixed surface is simply the condition that the quadratic singlet spurion and the quartic triplet spurion vanish.

This makes the structure of the RG flow transparent. At one loop, closure follows directly from the allowed $\SU(2)_H$ covariants. In the scalar-gauge sector the argument extends to all orders. Once the quartic triplet is absent, no triplet can be generated from symmetric products of the remaining quintet spurion, and the quadratic singlet then renormalizes only multiplicatively. The same mechanism is already visible in the complex-scalar toy model of Ref.~\cite{Haber:2025cbb}, where a missing spurionic charge protects a mass parameter, and the 2HDM result may be considered the generalization of this idea to the non-Abelian case.

General Yukawa interactions enlarge the set of available Higgs-family covariants and therefore spoil this scalar-sector protection. We consequently considered an exact symmetry $G$ imposed on the full 2HDM Lagrangian and studied the intersection $\mathcal S\cap G$. The resulting criterion is especially simple. If the real Higgs-family triplet contains no $G$-invariant direction, a triplet scalar counterterm cannot be generated. Among the six conventional 2HDM symmetry classes, $\SO(3)_H$, $\CP2$, and $\CP3$ satisfy this condition and protect the corresponding intersection with $\mathcal S$ to all perturbative orders. In contrast, $\mathbb Z_2$, $\U(1)_{PQ}$, and $\CP1$ retain $G$-invariant triplet directions, and explicit Yukawa covariants can populate those directions. These symmetries therefore do not, by themselves, protect the full surface.

More broadly, the spurion formulation shows that the RG stability of these relations in the 2HDM can be understood using conventional symmetry and representation theory. In the scalar-gauge sector, closure follows from the absence of the Higgs-family representations needed to regenerate the missing couplings. When Yukawa interactions are included, the same viewpoint gives a simple criterion for determining which exact symmetries of the full theory continue to protect the fixed relations.

\begin{acknowledgments}
The authors are supported by the National Science Foundation under grant number PHY-2514660 and by the Maryland Center for Fundamental Physics.
\end{acknowledgments}

\begin{appendices}
\renewcommand{\theequation}{\Alph{section}\arabic{equation}}
\setcounter{equation}{0}

\section{Character theory for $\SU(2)$}
\label{app:characters}

The material in this appendix is standard; see e.g.~\cite{Fulton2004,sternberg1994group} for detailed treatments and \cite{Dujava:2022vqz} for a review. We collect here the identities used in Sec.~\ref{sec:allorders}, spelling out the derivations in some detail since the techniques may be unfamiliar outside representation theory.

\subsection*{Characters and orthogonality}

Let $G$ be a compact Lie group and $(V,\rho)$ a finite-dimensional representation of $G$. Its character is the class function
\begin{equation}
  \chi_V(g) \equiv \Tr_V[\rho(g)].
\end{equation}
For $G=\SU(2)$, irreducible representations are labeled by spin $j=0,\tfrac12,1,\tfrac32,\ldots$, with dimension $2j+1$. A convenient representative of the conjugacy class is the element
\begin{equation}
  g(\theta)=
  \begin{pmatrix}
    e^{i\theta/2} & 0\\
    0 & e^{-i\theta/2}
  \end{pmatrix},
  \qquad 0\le \theta <2\pi.
\end{equation}
Any class function can therefore be regarded as a function of the single angle $\theta$. In the spin-$j$ representation, the weights are $m=-j,-j+1,\ldots,j$, so the character is
\begin{equation}
  \chi_j(\theta) = \sum_{m=-j}^{j} e^{im\theta} = \frac{\sin\!\left((2j+1)\theta/2\right)}{\sin(\theta/2)}.
\label{eq:SU2char}
\end{equation}

The invariant inner product on class functions is determined by the Haar measure. Using the Weyl integration formula for $\SU(2)$, one finds
\begin{equation}
  \langle f,h\rangle = \frac{1}{2\pi}\int_{0}^{2\pi} f(\theta)\,\overline{h(\theta)}\, 2\sin^2\!\frac{\theta}{2}\,d\theta.
\label{eq:Haar}
\end{equation}
For $\SU(2)$ the irreducible characters are real, so in all applications below the complex conjugation may be omitted. Character orthonormality then reads $\langle \chi_j,\chi_{j'}\rangle = \delta_{jj'}.$

If a representation $V$ decomposes as
\begin{equation}
  V = \bigoplus_j m_j\,V_j,
\end{equation}
then its character decomposes accordingly,
\begin{equation}
  \chi_V(\theta)=\sum_j m_j\,\chi_j(\theta),
\end{equation}
and orthonormality implies that the multiplicity of $V_{j'}$ is
\begin{equation}
  m_{j'}=\langle \chi_V,\chi_{j'}\rangle.
\label{eq:multiplicity}
\end{equation}
This is the identity used in the main text to extract the multiplicity of the triplet $\mathbf3$.

\subsection*{Characters of symmetric tensor powers}

We now derive the generating function for the characters of $\Sym^n(V)$. This is the only nontrivial identity needed in Sec.~\ref{sec:allorders}.

Let $V$ be a $d$-dimensional representation of $G$, and let the eigenvalues of $\rho(g)$ on $V$ be $x_1,\ldots,x_d$. For a fixed group element $g$, we may choose a basis $e_a$ of $V$ in which its action is diagonal,
\begin{equation}
    \rho(g)e_a=x_a e_a .
\end{equation}
The character $\chi_V(g)$ is simply the trace of this matrix, and hence $\chi_V(g)=\sum_a x_a$. Our goal is to determine the corresponding trace when $g$ acts on the symmetric tensor powers of $V$.

A convenient way to describe a basis of $\Sym^n(V)$ is by occupation numbers $(n_1,\ldots,n_d)$, where $n_a$ specifies how many times the basis vector $e_a$ appears in the symmetrized tensor. Thus
\begin{equation}
    n_a\ge0, \qquad \sum_{a=1}^d n_a = n.
\end{equation}
For example, a state with $(n_1,n_2,\ldots)=(2,1,0,\ldots)$ is the symmetrized version of $e_1\otimes e_1\otimes e_2$. Since $g$ acts independently on every factor, this state is an eigenvector of the induced action on $\Sym^n(V)$ with eigenvalue $x_1^{n_1}x_2^{n_2}\cdots x_d^{n_d}$. The character is the trace over all such states, so we sum these eigenvalues over every set of occupation numbers whose total is $n$,
\begin{equation}
  \chi_{\Sym^n(V)}(g) =
  \!\!\sum_{\substack{n_1,\ldots,n_d\ge0\\ n_1+\cdots+n_d=n}} x_1^{n_1}\cdots x_d^{n_d}.
\label{eq:Symnchar_explicit}
\end{equation}
Equivalently, $\chi_{\Sym^n(V)}(g)$ is the complete homogeneous symmetric polynomial of degree $n$ in the eigenvalues of $\rho(g)$.

It is useful to package the characters for all symmetric powers into a single generating function,
\begin{equation}
    Z_V(t,g) \equiv \sum_{n\ge0} t^n\,\chi_{\Sym^n(V)}(g).
\end{equation}
The auxiliary variable $t$ simply keeps track of the tensor degree with the coefficient of $t^n$ in $Z_V$ is the character of $\Sym^n(V)$. Substituting Eq.~\eqref{eq:Symnchar_explicit} gives
\begin{equation}
Z_V(t,g) = \sum_{n_1,\ldots,n_d\ge0}(tx_1)^{n_1}\cdots(tx_d)^{n_d}.
\end{equation}
Once the sum over all $n$ has been performed, the occupation numbers are independent and there is no longer a constraint fixing $n_1+\cdots+n_d$. The multiple sum therefore factorizes into $d$ ordinary geometric series,
\begin{align}
    Z_V(t,g) = \prod_{a=1}^{d} \left[\sum_{n_a\ge0}(tx_a)^{n_a}\right] = \prod_{a=1}^{d}\frac{1}{1-tx_a} \, .
\label{eq:symchar_product}
\end{align}
Thus the entire sequence of characters $\chi_{\Sym^n(V)}(g)$ is encoded in the simple product \eqref{eq:symchar_product} and expanding it in powers of $t$ recovers the character of each symmetric tensor power. For $|t|<1$ the series converges in the usual sense, although for our purposes it may equally well be viewed as a formal power series in $t$.

\subsection*{Absence of the triplet in $\Sym^{n}(\mathbf{5})$}

Specializing to the quintet $\mathbf{5}$ (spin $j=2$, weights $m=-2,-1,0,1,2$), the eigenvalues of $g(\theta)$ are $x_m=e^{im\theta}$. Substituting into Eq.~\eqref{eq:symchar_product} gives the generating function
\begin{align}
\label{eq:Z5}
  Z_{\mathbf{5}}(t,\theta) &\equiv \sum_{n\ge0} t^n\,\chi_{\Sym^n(\mathbf{5})}(\theta)
  \\
  &= \frac{1}{(1-t)(1-2t\cos\theta+t^2)(1-2t\cos2\theta+t^2)}.\nonumber
\end{align}
The multiplicity of $\mathbf{3}$ in $\Sym^n(\mathbf{5})$ is $m_{\mathbf{3}}^{(n)}=\langle\chi_{\Sym^n(\mathbf{5})},\chi_{\mathbf{3}}\rangle$ by Eq.~\eqref{eq:multiplicity}. Collecting these into a generating function and inserting Eqs.~\eqref{eq:Z5} and~\eqref{eq:Haar},
\begin{align}
  &\hspace{2cm}\mathcal{M}_{\mathbf{3}}(t) \equiv \sum_{n\ge0}m_{\mathbf{3}}^{(n)}\,t^n
  \nonumber \\
  &= \frac{1}{2\pi} \int_{0}^{2\pi} Z_{\mathbf{5}}(t,\theta)\, (1+2\cos\theta)\, 2\sin^2\!\tfrac{\theta}{2}\;d\theta,
\label{eq:M3integral}
\end{align}
where $\chi_{\mathbf{3}}(\theta)=1+2\cos\theta$.

Evaluating $\mathcal{M}_{\mathbf{3}}(t)$ as a contour integral over the unit circle, one finds that the coefficients are zero at every order in $t$, and hence
\begin{equation}
  \mathcal{M}_{\mathbf{3}}(t) = 0.
\end{equation}
Since a power series that vanishes identically has all coefficients zero, $m_{\mathbf{3}}^{(n)}=0$ for all $n\ge 0$, which proves Eq.~\eqref{eq:keyclaim}. This is also seen explicitly, for the first few terms we see
\begin{align}
  n=0:&\quad \langle\chi_{\mathbf{1}},\,\chi_{\mathbf{3}}\rangle = 0,
    \nonumber\\
  n=1:&\quad \langle\chi_{\mathbf{5}},\,\chi_{\mathbf{3}}\rangle = 0,
    \nonumber\\
  n=2:&\quad \langle\chi_{\mathbf{1}\oplus\mathbf{5}\oplus\mathbf{9}},\, \chi_{\mathbf{3}}\rangle = 0,
    \nonumber\\
  n=3:&\quad \langle\chi_{\mathbf{1}\oplus\mathbf{5}\oplus\mathbf{7}\oplus\mathbf{9}\oplus\mathbf{13}},\,\chi_{\mathbf{3}}\rangle = 0.
\end{align} 

\end{appendices}

\bibliographystyle{apsrev4-1}
\bibliography{Refs}
\end{document}